# What Is a Causal Effect When Firms Interact? Counterfactuals and Interdependence

January 2026

Mariluz Maté (mluz.mate@upct.es).

**Abstract**

Many empirical studies estimate causal effects in environments where economic units interact through spatial or network connections. In such settings, outcomes are jointly determined, and treatment-induced shocks propagate across economically connected units. A growing literature highlights identification challenges in these models and questions the causal interpretation of estimated spillovers. This paper argues that the problem is more fundamental. Under interdependence, causal effects are not uniquely defined objects—even when the interaction structure is correctly specified or consistently learned, and even under ideal identifying conditions. We develop a causal framework for firm-level economies in which interaction structures are unobserved but can be learned from predetermined characteristics. We show that learning the network, while necessary to model interdependence, is not sufficient for causal interpretation. Instead, causal conclusions hinge on explicit counterfactual assumptions governing how outcomes adjust following a treatment change. We formalize three economically meaningful counterfactual regimes—partial equilibrium, local interaction, and network-consistent equilibrium—and show that standard spatial autoregressive estimates map into distinct causal effects depending solely on the counterfactual adopted. We derive identification conditions for each regime and demonstrate that equilibrium causal effects require substantially stronger assumptions than direct or local effects. A Monte Carlo simulation illustrates that equilibrium and partial-equilibrium effects differ mechanically even before estimation, and that network feedback can strongly amplify bias when identifying assumptions fail. Taken together, our results clarify what existing spatial and network estimators can—and cannot—identify and provide practical guidance for empirical research in interdependent economic environments.

## 1. Introduction

A central challenge in empirical economics is to estimate causal effects in environments where economic activity is inherently interdependent across space or networks. In many settings of first-order policy and economic interest—such as firm-level interventions that affect competitors or suppliers, place-based policies with geographically diffuse impacts, or the propagation of technological and demand shocks through production networks—outcomes for a given unit depend not only on its own treatment status, but also on the treatments and outcomes of other economically connected units (Aghion et al. 2016; De Loecker et al. 2020; Kline and Moretti 2014; Busso et al. 2013; Acemoglu et al. 2012; Carvalho 2014). As a result, treatments typically generate spillovers and feedback effects that extend beyond directly treated units.

This interdependence fundamentally complicates causal inference. Standard frameworks based on isolated units and well-defined potential outcomes no longer apply directly: interference across units violates the stable unit treatment value assumption, blurring both the definition and identification of causal effects (Manski 1993; Hudgens and Halloran 2008). Even when interference is restricted or approximated, causal conclusions remain sensitive to assumptions about how treatment effects propagate across units. Leung (2022), for instance, formalizes causal inference under approximate neighborhood interference and shows that identification depends critically on specifying which interactions are allowed to matter and which are excluded.

A large empirical literature has responded to these challenges by explicitly modeling interdependence through spatial or network econometric specifications. Spatial autoregressive (SAR) and related models are now routinely employed to capture spillovers across firms, regions, or locations (Anselin 1988; LeSage and Pace 2009). In applied work, estimated spatial dependence parameters are often interpreted as causal spillover effects, implicitly assuming that the econometric model accurately captures how treatments propagate through space. At the same time, this practice has attracted sustained criticism. Gibbons and Overman (2012) question whether many applications of spatial econometrics deliver meaningful causal insights at all, while Debarsy and Le Gallo (2025) emphasize that both identification and causal interpretation in spatial models hinge on strong—and often implicit—assumptions that are difficult to justify empirically.

Two distinct but closely related issues underlie these concerns. The first concerns the specification of the interaction structure itself. Most empirical applications impose a spatial weights matrix ex ante, typically based on geographic distance or contiguity. While convenient, this approach assumes that economic interactions are known and exogenous. This assumption is increasingly problematic in firm-level environments, where proximity is multidimensional—geographic, technological, input–output, or informational—and interaction patterns are only imperfectly observed. Recent contributions address this limitation by proposing methods to estimate interaction structures directly from predetermined characteristics (Qu, Lee, and Yang 2021; Merk and Otto 2022; Gao and Ding 2025), offering more flexible and empirically grounded representations of economic connectivity.

The second issue—which is the focus of this paper—concerns the definition of the causal effect itself. In the standard potential-outcomes framework, causal effects are defined by comparing outcomes across hypothetical worlds in which only the treatment assignment changes (Rubin 1974; Imbens and Rubin 2015). Such comparisons rely on a well-defined counterfactual: what would have happened to a unit had it received a different treatment, holding all else constant. In environments characterized by spatial or network interdependence, this notion becomes inherently ambiguous. Changing the treatment status of one unit typically affects the outcomes of other units, which may then feed back into the original unit. As a result, it is no longer clear what should be held fixed—and what should be allowed to adjust—when defining the counterfactual outcome.

A central lesson of modern causal inference is that causal effects are defined relative to explicit counterfactual experiments. As emphasized by Heckman and Vytlacil (2007), the interpretation of any treatment effect hinges on the underlying policy experiment and on which margins of adjustment are allowed to vary. Different counterfactual experiments correspond to different causal parameters, even when the same data-generating process is considered. Recent syntheses reinforce this point. Imbens (2024) stresses that identification and estimation are secondary to a prior conceptual step: clarifying which hypothetical intervention and adjustment process define the causal object of interest.

This ambiguity is not merely conceptual. A growing literature on causal inference under interference shows that different assumptions about how interference operates lead to different causal effects, even when the same data and estimation strategy are used.

Randomization-based approaches highlight the central role of exposure mappings in linking treatment assignments to potential outcomes under interference, demonstrating that the choice of exposure mapping fundamentally shapes the causal effects that can be defined and estimated (Aronow and Samii 2017). In networked settings, causal inference therefore requires an explicit specification of how treatment assignments map into potential outcomes through the interaction structure. As shown by Athey et al. (2018), different interference structures and exposure mappings can yield distinct causal effects even when the estimator and data are held fixed. Relatedly, statistical approaches to causal inference under unknown or arbitrary interference show that standard estimators target different causal contrasts depending on how spillovers are summarized and represented, and that misspecification of these structures can fundamentally alter causal conclusions (Sävje et al., 2021). More broadly, a large literature on peer and network effects documents the pervasiveness of interdependence across economic environments (Bramoullé et al., 2009; Bramoullé et al. 2020).

This paper brings these strands together. We argue that causal inference under endogenous spatial interdependence requires both learning the interaction structure and explicitly redefining the counterfactual. Learning the network is necessary to avoid misspecification and spurious spillovers, but it is not sufficient for causal interpretation. Even with a correctly specified or consistently learned interaction structure, causal effects remain ill-defined unless the researcher specifies how shocks propagate through the network in the counterfactual scenario. We develop a causal framework for firm-level economies in which interaction structures are unobserved but can be learned from predetermined characteristics. Within this framework, we show that different counterfactual assumptions—such as holding other units' outcomes fixed, allowing only local responses, or allowing full network feedback—correspond to distinct and economically meaningful causal effects. Importantly, we demonstrate that commonly estimated parameters in spatial autoregressive models map to different causal objects depending on these counterfactual assumptions, a distinction that is rarely acknowledged in applied work.

Our contribution is threefold. First, we clarify that under endogenous interdependence there is no single causal effect of a treatment, but rather a family of effects indexed by explicit counterfactual regimes governing how interactions adjust to treatment changes.

Second, we derive identification conditions linking each counterfactual regime to specific assumptions on treatment assignment and interdependence, thereby clarifying what spatial econometric models can—and cannot—identify. Third, we show that a large share of the empirical literature implicitly relies on undefined or internally inconsistent counterfactuals when interpreting spillover effects, extending recent critiques of spatial identification to settings with endogenous interaction structures. By placing the counterfactual at the center of causal analysis under interdependence, this paper provides practical guidance for empirical research in spatial and networked economic environments.

The remainder of the paper proceeds as follows. Section 2 introduces a motivating example illustrating why causal effects are ambiguous under interdependence. Section 3 presents the framework and formalizes endogenous interaction structures. Section 4 defines alternative counterfactual regimes and their economic interpretation. Section 5 discusses identification, and Section 6 presents the estimation strategy and simulation evidence.

## 2. A Motivating Example: Firm-Level Treatments under Economic Interdependence

This section illustrates why causal effects are inherently ambiguous in economic environments characterized by local interactions, even when the interaction structure is correctly specified or consistently estimated. The objective is not to introduce new econometric techniques, but to clarify the definition of the causal object itself. We show that, once interdependence is acknowledged, causal effects cannot be defined without an explicit statement of the counterfactual regime governing how shocks propagate through the economy.

With this purpose, let us consider an economy composed by a finite number of firms indexed by $i = 1, \ldots, N$. Firms are located in space and interact locally through competition, supply-chain relationships, labor market linkages, or knowledge spillovers. Such interactions are central to many applied contexts, including industrial policy, place-based interventions, and innovation diffusion (Aghion et al., 2016; Kline and Moretti, 2014; Acemoglu et al., 2012; Carvalho, 2014). Let $D_i \in \{0,1\}$ denote a binary treatment—such as a subsidy, tax credit, or productivity-enhancing policy—assigned to a firm $i$, and let $Y_i$ denote an outcome of interest, such as productivity, output, or employment. In a hypothetical environment without interdependence, the causal effect of the treatment on firm $i$ would be defined in the standard way as the difference between potential outcomes,

$$Y_i(1) - Y_i(0) \qquad (1)$$

where $Y_i(d)$ denotes the outcome that would be observed for firm $i$ under treatment status $d$, holding all other aspects of the environment fixed. This definition relies on the implicit assumption that changing the treatment of firm $i$ does not affect the outcomes of other firms. Under this assumption, causal effects are well defined and can be identified under standard conditions (Rubin, 1974; Imbens and Rubin, 2015).

Now suppose instead that firms interact locally. Let $N_i$ denote the generally unobserved set of firms whose outcomes affect a firm $i$. Outcomes are generated according to (2)

$$Y_i = f(D_i, \{Y_j : j \in N_i\}, X_i, \varepsilon_i) \qquad (2)$$

where $X_i$ are predetermined firm characteristics and $\varepsilon_i$ is an idiosyncratic shock. This formulation captures the idea that firm outcomes depend not only on own treatment status, but also on the outcomes of economically connected firms, as emphasized in the literature on social interactions and spatial dependence (Manski, 1993; Anselin, 1988). In this context, changing the treatment status of firm $i$ generally affects the outcomes of firms in $N_i$, which in turn feed back into $Y_i$. As a result, the potential outcome $Y_i(d)$ is no longer well defined unless one specifies how other firms' outcomes are allowed to adjust in response to the treatment change in $i$. The standard potential-outcomes notation masks this issue by implicitly assuming that changes in one unit's treatment do not affect the outcomes of others. This is precisely the starting point of the modern interference literature: under interdependence, causal effects are inseparable from the counterfactual assumptions that restrict or summarize how shocks propagate across units (Hudgens and Halloran, 2008; Sävje et al., 2021; Athey et al., 2018). In spatial settings, related points arise when causal parameters depend on how spillovers are mapped into exposure measures (Qiu and Tong, 2021).

To see the nature of the problem more clearly, consider the causal effect of treating a single firm $i$. *What does it mean to compare the observed outcome under treatment to a* *counterfactual outcome without treatment?* In the presence of interdependence between companies, this comparison is not unique. Different counterfactual regimes correspond to different answers to the question of what should be held fixed and what should be allowed to adjust when treatment status changes. One possibility is a **partial-equilibrium counterfactual**, in which the treatment status of firm $i$ changes while the outcomes of all other firms are held fixed at their pre-treatment levels. This counterfactual isolates the direct effect of the treatment on the treated firm, abstracting entirely from spillovers and feedback effects. Economically, it corresponds to an experiment in which the treated firm adjusts its behavior in response to the policy, but its competitors, suppliers, and customers do not respond at all. For example, when evaluating a subsidy to a manufacturing firm, the partial-equilibrium effect asks how the firm's output or productivity would change if it received the subsidy, holding prices, competitors' outputs, and input availability fixed. This is the type of causal effect often implicitly targeted in empirical work, even when spatial dependence is present, because it aligns closely with standard regression interpretations and requires relatively weak identifying assumptions.

A second possibility allows for a **local-interaction counterfactual**, in which firms that interact directly with firm $i$ are allowed to respond to the treatment according to the underlying economic mechanism, while the rest of the economy is held fixed. This counterfactual captures first-order spillovers but rules out broader equilibrium adjustments and higher-order feedback effects. Continuing the subsidy example, this counterfactual would allow nearby competitors to adjust their output or prices in response to the treated firm's expansion, or immediate suppliers to adjust quantities, while assuming that more distant firms and markets remain unaffected. This regime is often implicitly invoked when empirical studies interpret spatial lag terms as localized spillover effects. However, it requires a substantive assumption that the economic effects of the treatment dissipate quickly and do not propagate beyond the immediate neighborhood, an assumption that may be difficult to justify in dense or highly connected economic environments.

A third possibility is a **network-consistent (equilibrium) counterfactual**, in which the treatment-induced shock is allowed to propagate through the entire network of interactions until a new equilibrium is reached. Under this regime, all firms connected—directly or indirectly—to firm $i$ adjust their outcomes, and these adjustments feed back into one another through the interaction structure. In the subsidy example, this counterfactual allows not only direct competitors and suppliers to respond, but also second- and third-order effects: competitors' competitors adjust, prices change in related markets, and these adjustments ultimately feed back into the treated firm itself. This counterfactual captures the total equilibrium effect of the treatment, including all indirect and feedback effects, and is most relevant for policy questions concerned with market-wide reallocation, aggregate productivity, or welfare. At the same time, it is the most demanding in terms of identifying assumptions, as it requires ruling out correlation between treatment assignment and unobserved shocks anywhere in the system.

Together, these counterfactuals illustrate that there is no single causal effect of treating firm $i$ once interdependence is acknowledged. Each counterfactual corresponds to a distinct economic question and a distinct interpretation of empirical estimates. Making these distinctions explicit is essential for aligning empirical strategies with the policy questions they are intended to answer. However, these distinctions are rarely made explicit in empirical practice. As a result, empirical estimates are often interpreted as measuring "spillover effects" without a clear statement of the counterfactual world to which they correspond.

To connect this discussion to standard empirical models, suppose that the interaction structure can be summarized by a weighted network matrix $W$, where the elements $w_{ij} > 0$ if firms $j$ and $i$ are connected, and $w_{ij} = 0$ otherwise. Outcomes are generated by the spatial autoregressive equation (3)

$$Y = \rho W Y + \beta D + X \gamma + \varepsilon \qquad (3)$$

as in spatial autoregressive (SAR) and related models widely used in applied work (Anselin, 1988; LeSage and Pace, 2009). Recent advances in the literature allow researchers to estimate $W$ from predetermined firm characteristics rather than imposing it ex ante (Gao and Ding, 2025; Merk and Otto, 2022; Qu et al., 2021). Doing so addresses an important source of misspecification and brings empirical models closer to economic reality. However, even if the interaction matrix $W$ is correctly learned, the causal interpretation of the parameters $\beta$ and $\rho$ remains ambiguous. Estimating $W$ informs us about how outcomes are interconnected in the observed data, but it does not determine how these interconnections should be treated in the counterfactual comparison. Holding outcomes of other firms fixed while changing treatment corresponds to one counterfactual regime. Allowing outcomes to adjust through the interaction matrix corresponds to another. Allowing full feedback and equilibrium adjustment corresponds to yet another. These counterfactual regimes lead to different causal objects, even though they rely on the same estimated interaction structure.

The key implication is that, in environments characterized by endogenous interdependence, there is no single causal effect of a treatment. Instead, causal effects are indexed by counterfactual assumptions about how interactions operate when treatment status changes. Empirical estimates that do not explicitly state which counterfactual regime they correspond to implicitly mix these concepts. Consistent with Debarsy and Le Gallo (2025), this ambiguity can generate misleading causal interpretations even when identification is otherwise credible. The remainder of this paper builds on this insight by developing a framework that makes counterfactual assumptions explicit, clarifies their identifying requirements and maps common spatial econometrics objects into well-defined causal effects.

## 3. Conceptual Framework

This section develops a conceptual framework to formalize causal analysis in environments where economic units interact through an endogenous structure of interdependence. The objective is to clarify how causal effects can be defined and interpreted when outcomes are jointly determined across firms, and to make explicit the additional structure required to move from general interference to empirically meaningful causal objects.

Consider a finite population of firms indexed by $i = 1, \ldots, N$. Each firm is characterized by a vector of predetermined characteristics $X_i$ and is subject to a binary treatment $D_i \in \{0,1\}$. $D = (D_1, \ldots, D_N)$ denote the vector of treatment assignments. For each firm $i$, let $Y_i(D)$ denote the potential outcome under treatment assignment $D$. We allow for unrestricted interdependence across firms: the outcome of firm $i$ may depend not only on its own treatment status but also on the treatment assignments of other firms. Unlike frameworks that impose partial interactions or predefined exposure mappings, we do not assume that the set of economically relevant interconnections is known ex ante. Instead, we begin from a fully general representation in which potential outcomes depend on the entire treatment vector,

$$Y_i(D) = g_i(D_i, D_{-i}, X, \varepsilon_i) \qquad (3)$$

where $g_i(\cdot)$ is an unknown structural outcome function, $D_{-i}$ denotes the treatment assignments of all firms other than $i$, $X = (X_1, \ldots, X_N)$ collects predetermined characteristics, and $\varepsilon_i$ is an idiosyncratic shock. This formulation highlights the core difficulty posed by interdependence: absent further restrictions, the number of potential outcomes grows exponentially with the number of firms, rendering imprecise causal effects. Defining and interpreting causal effects therefore requires additional structure on how economic interactions operate.

To discipline this general interaction setting, we summarize interdependence through an interaction matrix $W \in R^{N \times N}$, where the elements $w_{ij} \geq 0$ captures the strength of the interaction between the firms $j$ and $i$, and $w_{ii} = 0$. Crucially, we do not treat the interaction structure as known or exogenous. Instead, we assume that it is generated by a mapping

$$W = W(\theta; X) \qquad (4)$$

where θ is a finite-dimensional parameter vector and $X$ consists of predetermined firm characteristics. This formulation reflects the idea that economic proximity—whether geographic, technological, input–output, or informational—is latent and must be inferred from observables. A central insight in the network literature is that the structure of interactions can be exploited to discipline interdependence and identify interaction parameters. Bramoullé et al. (2009) show that network topology can be used to identify peer effects even when outcomes are jointly determined. We build on this insight but emphasize a distinct point: even when the interaction structure is known or consistently learned, causal effects remain unclear unless the counterfactual adjustment regime is made explicit.

By construction, the interaction structure is predetermined with respect to treatment assignment, rejecting simultaneity between treatment and network formation. This assumption aligns with recent work that estimates spatial or network weights from predetermined characteristics rather than imposing them ex ante (Qu et al., 2021; Merk and Otto 2022; Gao and Ding 2025). Conditional on the interaction structure, outcomes are jointly determined according to the spatial autoregressive equilibrium condition,

$$Y = \rho WY + \beta D + X\gamma + \varepsilon \tag{5}$$

where $Y = (Y_1, \ldots, Y_N)^\top$ and $\varepsilon = (\varepsilon_1, \ldots, \varepsilon_N)^\top$. Solving this system yields the equilibrium mapping

$$Y(D) = (I - \rho W)^{-1}(\beta D + X\gamma + \varepsilon) \tag{6}$$

This expression makes explicit that potential outcomes depend on the entire treatment vector through the interaction structure. A change in the treatment status of a single firm generally affects the outcomes of other firms and feeds back through the network, so that the potential outcome $Y_i(D)$ cannot be associated with a scalar treatment status $D_i$ alone. Equation (6) plays a central role in what follows. It clarifies how treatment-induced shocks propagate through the interaction structure and highlights the distinction between model parameters—such as β and ρ—and the causal objects they may or may not represent.

Learning the interaction structure from data is therefore a necessary step to model interdependence and avoid misspecification. However, it is not sufficient for causal interpretation. Even with a correctly specified or consistently learned interaction structure, the causal effect of a treatment remains ambiguous unless one specifies the counterfactual rule governing which components of the system are held fixed and which are allowed to adjust when treatment status changes. Without such a counterfactual specification, causal effects are not uniquely defined, even when the network is known.

**4. Counterfactuals under Endogenous Spatial Interdependence**

When outcomes are jointly determined through an interaction structure, causal effects cannot be defined independently of assumptions about how the system adjusts following a treatment change. As established in Section 3, potential outcomes are indexed by the entire treatment vector, and a change in the treatment status of a single firm generally propagates through the network of economic interactions. Defining a causal effect therefore requires an explicit counterfactual rule specifying which components of the system are held fixed and which are allowed to adjust.

This observation can be stated more formally. Consider an outcome system characterized by interdependence and summarized by the reduced-form relationship (5) where the interaction structure W is known or consistently learned from predetermined characteristics. For a given change in the treatment status of a single unit iii, there exist multiple economically reasonable counterfactual rules governing how the outcomes of other units adjust. Each rule induces a distinct causal effect for unit iii, even though the underlying model, parameters $(\beta, \rho)$, and interaction structure W are held fixed. Consequently, under interdependence, causal effects are not uniquely defined objects unless the counterfactual adjustment regime is explicitly specified. This point resonates with a broader insight in modern causal inference: causal effects are defined relative to hypothetical interventions, not model parameters alone (Heckman and Vytlacil, 2007; Imbens, 2024).

To clarify the nature of the problem, consider a policy intervention that changes the treatment status of a single firm $i$ from untreated to treated, holding the treatment status of all other firms fixed. In the absence of interdependence, the causal effect would be well defined as the difference between two scalar potential outcomes. Under interdependence, however, this comparison is incomplete. Changing $D_i$ typically affects not only the outcome of firm i directly, but also the outcomes of other firms through economic interactions, which may in turn feed back into $Y_i$. Different assumptions about how these adjustments unfold correspond to different causal effects, even when the underlying structural model is the same. This is precisely the problem highlighted in the interference literature, where causal objects depend on how exposure to others' treatments is defined **(**Hudgens and Halloran, 2008; Sävje et al., 2021**).**

A first counterfactual isolates the direct effect of treatment by holding the outcomes of all other firms fixed at their pre-treatment levels. Under this partial-equilibrium counterfactual, the causal effect for firm $i$ is defined as

$$\Delta_i^{PE} = Y_i(D_i = 1, Y_{-i} = Y_{-i}^0) - Y_i(D_i = 0, Y_{-i} = Y_{-i}^0). \qquad (7)$$

where $Y^0_{-i}$ denotes the vector of outcomes in the absence of the treatment. Economically, this counterfactual corresponds to an experiment in which firm $i$ adjusts its behavior in response to the policy while competitors, suppliers, and customers do not respond at all. It captures a purely direct effect, abstracting entirely from spillovers and feedback effects. While this counterfactual is often implausible in environments characterized by strategic interaction or market competition, it remains relevant as a benchmark and aligns closely with the causal interpretation implicitly adopted in much empirical work based on regression models.

A second counterfactual relaxes this assumption by allowing firms that interact directly with the treated firm to respond to the treatment, while discarding broader equilibrium adjustments. Under this local-interaction counterfactual, the treatment of the firm I affects the outcomes of firms directly connected to I through the interaction structure, but higher order feedback effects are not permitted. Formally, the causal effect is defined as:

$$\Delta^{LI}_i = Y_i(D_i = 1, Y_{N_i} = Y^1_{N_i}, Y_{-N_i} = Y^0_{-N_i}) - Y_i(D_i = 0, Y_{N_i} = Y^0_{N_i}, Y_{-N_i} = Y^0_{-N_i}). \quad (8)$$

where $N_i$ denotes the set of firms directly connected to $i$. This counterfactual captures first-order spillovers without allowing for recursive feedback through the network. It corresponds closely to empirical practices that interpret spatial lag terms as localized spillover effects (Bramoullé et al., 2009; Bramoullé et al., 2020). However, it relies on a substantive assumption that the economic effects of the treatment dissipate quickly and do not propagate beyond the immediate neighborhood—an assumption that may be difficult to justify in dense or highly connected economic environments.

A third counterfactual allows the treatment-induced shock to propagate fully through the interaction structure until a new equilibrium is reached. Under this network consistent counterfactual, the causal effect of treating firm I is defined as the difference between equilibrium outcomes under alternative treatment vectors,

$$\Delta^{NC}_i = Y_i(D_i = 1, D_{-i} = 0) - Y_i(D_i = 0, D_{-i} = 0). \quad (9)$$

where both outcomes are evaluated at their respective equilibrium values implied by the model. Using the equilibrium mapping introduced in Section 3, this effect can be written as:

$$\Delta_i^{NC} = e_i^\top (I - \rho W)^{-1} \beta e_i. \tag{10}$$

where $e_i$ is a selection vector with a one in position $i$. This counterfactual captures the total equilibrium effect of the treatment, including all indirect and feedback effects transmitted through the interaction structure. It is the relevant causal object for policy evaluation in settings where spillovers, strategic responses and general equilibrium adjustments are central to the economic mechanisms. At the same time, it is the most demanding in terms of identifying assumptions, as it requires ruling out correlation between treatment assignment and unobserved shocks anywhere in the network.

These counterfactual regimes correspond to distinct and economically meaningful causal effects. None is inherently more correct than the others. The appropriate counterfactual depends on the policy question under consideration. A policymaker interested in the immediate response of a treated firm may focus on a partial-equilibrium effect, while one concerned with competition, reallocation, or aggregate outcomes may care about equilibrium effects. Crucially, however, these counterfactuals are not interchangeable. The same estimated model parameters can correspond to different causal effects depending on the counterfactual regime adopted. Without an explicit statement of the counterfactual world under consideration, empirical results lack a well-defined causal interpretation —an issue increasingly emphasized in recent critiques of spatial causal inference (Debarsy and Le Gallo, 2025).

**5. Identification of Causal Effects under Endogenous Interdependence**

The counterfactual regimes introduced in the previous section define distinct causal effects. Whether these effects can be learned from observed data depends on the assumptions governing treatment assignment and on how interdependence enters the system. Identification is therefore inseparable from the choice of counterfactual. Different causal effects require different identifying assumptions, a distinction that is often blurred in empirical applications of spatial and network models (Mansky, 1993; Debarsy and Le Gallo, 2025). In this section, we maintain the assumption that the interaction structure is predetermined with respect to treatment assignment.

**Assumption 1 (Predetermined Interaction Structure).** The interaction matrix $W = W(\theta; X)$ is a deterministic function of predetermined firm characteristics $X$. Conditional on $X$, the interaction structure is fixed with respect to the treatment assignment $D$.

This assumption rules out simultaneity between treatment assignment and the formation of economic connections. It is consistent with environments where economic proximity is determined by slow-moving characteristics such as location, technology, or organizational structure, while treatments are assigned after these characteristics are realized. At the same time, it allows interaction patterns to be endogenous to the economic environment rather than imposed ex ante, as in standard spatial econometric practice (Anselin, 1988; LeSage and Pace, 2009).

We begin with the identification of the partial-equilibrium counterfactual. Because this counterfactual holds the outcomes of all other firms fixed by construction, it eliminates any role for spillovers or feedback effects by construction. Identification therefore relies on conditions analogous to those used in settings without interference.

**Proposition 1 (Identification of Partial-Equilibrium Effects)[1].** Under Assumption 1 and conditional exogeneity of treatment assignment,

$$D_i \perp \varepsilon_i \mid X_i. \qquad (11)$$

the partial-equilibrium effect $\Delta_i^{PE}$ is identified by the direct treatment coefficient β. This result highlights a basic but important point. Partial-equilibrium effects are identifiable under relatively weak assumptions, precisely because they ignore economically meaningful interactions and may substantially understate policy impacts in environments characterized by interdependence. As a consequence, they may substantially understate policy impacts in environments where interactions, spillovers, or strategic responses are economically important (Heckman and Vytlacil, 2007).

Identification becomes more demanding once local spillovers are allowed. Under the local-interaction counterfactual, the treatment of firm i affects the outcomes of firms directly connected to it, which in turn influence $Y_i$. Identification therefore requires that treatment assignment be uncorrelated not only with the treated firm's idiosyncratic shock, but also with shocks affecting its immediate neighbors.

---

[1] Formal proofs of Propositions 1–3 are developed in Appendix A. These proofs formalize the identification arguments underlying each counterfactual regime and show explicitly how the strength of the identifying assumptions increases with the scope of equilibrium adjustment allowed under the counterfactual.

**Assumption 2 (Local Exogeneity).** For all firms $j$ such that $w_{ij} > 0$,

$$D_i \perp \varepsilon_j \mid X. \tag{12}$$

Before stating the formal identification result, it is useful to clarify how Proposition 2 fits into the broader logic of this section. While identification of partial-equilibrium effects requires only individual-level exogeneity, identification under interdependence depends critically on which responses are allowed in the counterfactual. The local-interaction counterfactual occupies an intermediate position. It permits direct spillovers to economically connected neighbors, but deliberately excludes higher-order feedback and equilibrium adjustments. As a result, Proposition 2 does not establish identification of a total or equilibrium effect. Instead, it establishes identification of *first-order causal spillovers*, the object implicitly targeted when spatial lag terms are interpreted as localized spillover effects in empirical work (Bramoullé et al., 2020). Making this distinction explicit is essential for correctly interpreting spatial coefficients.

**Proposition 2 (Identification of Local-Interaction Effects).** Under Assumptions 1 and 2, the local-interaction effect $\Delta_i^{LI}$ is identified by the first-order spatial propagation of the treatment through the interaction structure, $W$, holding higher-order feedback effects fixed.

This identification strategy aligns with empirical approaches that interpret spatial lag coefficients as localized spillover effects. However, it relies on a substantive assumption that higher-order interactions are either negligible or deliberately excluded. In dense networks or in environments with strong equilibrium feedback, this assumption may be difficult to justify empirically (Leung, 2022).

The strongest identifying requirements arise for the network-consistent counterfactual. When the treatment-induced shock is allowed to propagate fully through the interaction structure, equilibrium feedback amplifies the consequences of any correlation between treatment assignment and unobserved shocks. Identification of the equilibrium effect therefore hinges on global exogeneity.

**Assumption 3 (Global Exogeneity).** The treatment assignment vector $D$ is conditionally independent of the vector of idiosyncratic shocks $\varepsilon$ given $X$:

$$D \perp \varepsilon \mid X \tag{13}$$

**Proposition 3 (Identification of Network-Consistent Effects).** Under Assumptions 1 and 3, the network-consistent effect $\Delta_i^{NC}$ is identified by the equilibrium mapping implied by the reduced form model.

This result establishes that full-equilibrium causal effects are identifiable only under strong conditions that exclude correlation between treatment assignment and unobserved shocks anywhere in the network. Such assumptions may be plausible in experimental or carefully designed quasi-experimental settings, but they are rarely satisfied in observational data without explicit justification (Hudgens and Halloran, 2008; Sävje, et al., 2021).

Taken together, these propositions clarify the relationship between counterfactual choice and identification. Estimating a spatial or network model does not, by itself, determine which causal effect is being identified. Partial-equilibrium, local-interaction, and network-consistent effects correspond to different counterfactual regimes and require progressively stronger identifying assumptions (summarized in Table 1). Empirical studies that do not make these distinctions explicit risk attributing causal meaning to parameters that are only descriptively valid.

Our contribution is complementary to existing critiques of spatial causal inference. Debarsy and Le Gallo (2025) emphasize that identification of spatial effects hinges on strong and often implicit assumptions. In contrast, we show that the problem is more fundamental: under interdependence, even with a correctly specified or learned interaction structure and under ideal identifying conditions, causal effects are not uniquely defined objects. They are indexed by explicit counterfactual regimes governing how interactions adjust to treatment changes. By making these regimes explicit and linking them to identification requirements, our framework clarifies what existing spatial estimators can—and cannot—identify.

| Table 1- Counterfactual regimes and identification under interdependence ||||||
| Counterfactual regime | What varies when ($D_i$) changes | What is held fixed | Causal object identified | Required exogeneity | Typical empirical interpretation |
| --- | --- | --- | --- | --- | --- |
| Partial equilibrium (PE) | Outcome of unit (i) only | Outcomes of all other units | Direct (own) effect | Individual exogeneity ($D_i \perp \varepsilon_i \mid X_i$) | Standard regression coefficient |
| Local interaction (LI) | Unit (i) and its direct neighbors | Higher-order feedback | First-order spillovers | Local exogeneity toward neighbors | "Local" spatial spillovers |
| Network-consistent (NC) | All units through equilibrium | Nothing | Total equilibrium effect | Global exogeneity ($D \perp \varepsilon \mid X$) | SAR impacts / spatial multipliers |

**6. Monte Carlo Simulation: Counterfactuals, Networks, and Causal Interpretation**

This section presents a Monte Carlo simulation designed to make the paper's core argument transparent. In environments characterized by spatial or network interdependence, causal conclusions depend fundamentally on the counterfactual regime under consideration, even when the econometric model is correctly specified, and treatment assignment is exogenous. The simulation is deliberately modest. We do not seek to propose a novel estimator, not to replicate empirically rich settings. Rather, the simulation is designed as a controlled experiment that isolates the role of counterfactual assumptions in shaping causal interpretation.

The data-generating process features a finite population of firms interacting through a sparse network $W^\star$, constructed from predetermined geographic and economic characteristics. Outcomes follow a spatial autoregressive structure with a homogeneous direct treatment effect $\beta = 1$ and spatial dependence parameter $\rho = 0.4$. This setting is intentionally conventional and transparent, closely reflecting the specifications commonly used in applied spatial econometrics (Anselin, 1988; LeSage and Pace, 2009). As a result, any differences that emerge in the analysis can be attributed unambiguously to counterfactual interpretation, rather than to modelling complexity or estimation failure.

Before turning to estimation, we characterize the causal effects implied by the data generating process itself. By construction, the partial-equilibrium causal effect equals the direct treatment coefficient, $\Delta^{PE} = \beta = 1$. In contrast, the network-consistent (equilibrium) causal effect allows treatment-induced shocks to propagate through the interaction structure and feed back to the treated firm. At the firm level, this effect is given by

$$\Delta_i^{NC} = \beta \cdot [\,(I - \rho\, W^\star)^{-1}\,]_{ii} \qquad (13)$$

Averaging across firms yields an equilibrium effect of approximately 1.037, corresponding to an amplification of about 3.7 percent relative to the partial-equilibrium effect. This difference arises mechanically from network feedback and is present even in the absence of estimation error. Table 2 reports these true counterfactual effects.

| Table 2. True counterfactual effects implied by the DGP | |
|---|---|
| Quantity | Value |
| $\Delta^{PE}$ | 1.000 |
| $\Delta^{NC}(average)$ | 1.037 |
| $\Delta^{NC}/\Delta^{PE}$ | 1.037 |

Although the direct effect is homogeneous by construction, equilibrium effects are heterogeneous across firms. Figure 1 shows the distribution of the amplification factor $\Delta_i^{NC}/\Delta^{PE}$, which reflects only firms' positions within the interaction network. This heterogeneity is entirely structural and does not depend on any econometric procedure or sampling variation.

**Figure 1. Distribution of true amplification: $\Delta_i^{NC}/\Delta^{PE}$**

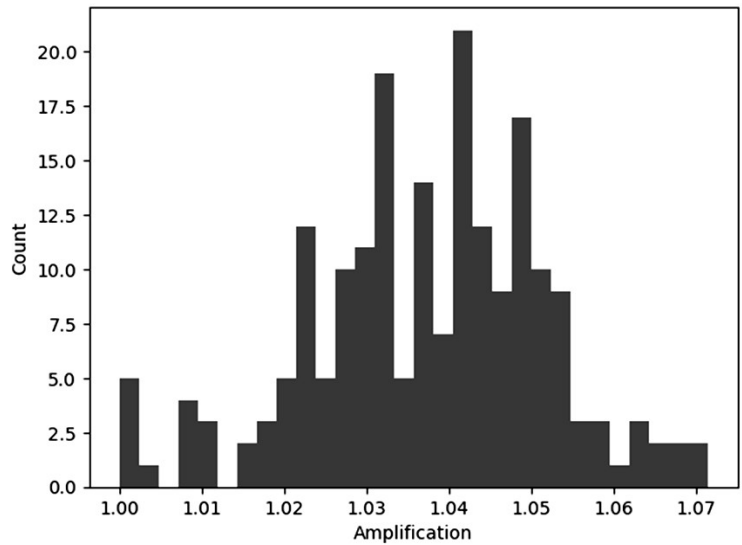

We next turn to estimation. In each Monte Carlo replication, treatment is assigned exogenously, outcomes are generated from the SAR DGP, and a spatial autoregressive model is estimated using the true interaction matrix $W^\star$. From the estimated parameters $(\hat{\beta}, \hat{\rho})$, we recover both the estimated direct effect $\hat{\beta}$ and the implied network-consistent effect $\hat{\Delta}^{NC}$ by mapping the estimates into the equilibrium expression. Table 3 reports bias, dispersion, and root mean squared error for these quantities.

| Table 3. Monte Carlo performance under exogenous assignment | | | |
|---|---|---|---|
| Estimator | True | Mean | Bias |
| $\hat{\beta}$ | 1.0 | 0.968 | −0.032 |
| $\hat{\rho}$ | 0.4 | 0.703 | +0.303 |
| $\hat{\Delta}^{NC}$ | 1.037 | 1.126 | +0.089 |

Two patterns are immediate. First, the estimated treatment coefficient $\hat{\beta}$ is centered close to the true partial-equilibrium effect, indicating that the estimator performs reasonably well for the direct-effect object under exogenous assignment. Second, the implied equilibrium effect $\hat{\Delta}^{NC}$ exhibits substantially larger bias and dispersion. This difference does not reflect a failure of estimation. It arises because equilibrium counterfactuals combine estimation error in both $\hat{\beta}$ and $\hat{\rho}$ with feedback effects embedded in the interaction structure.

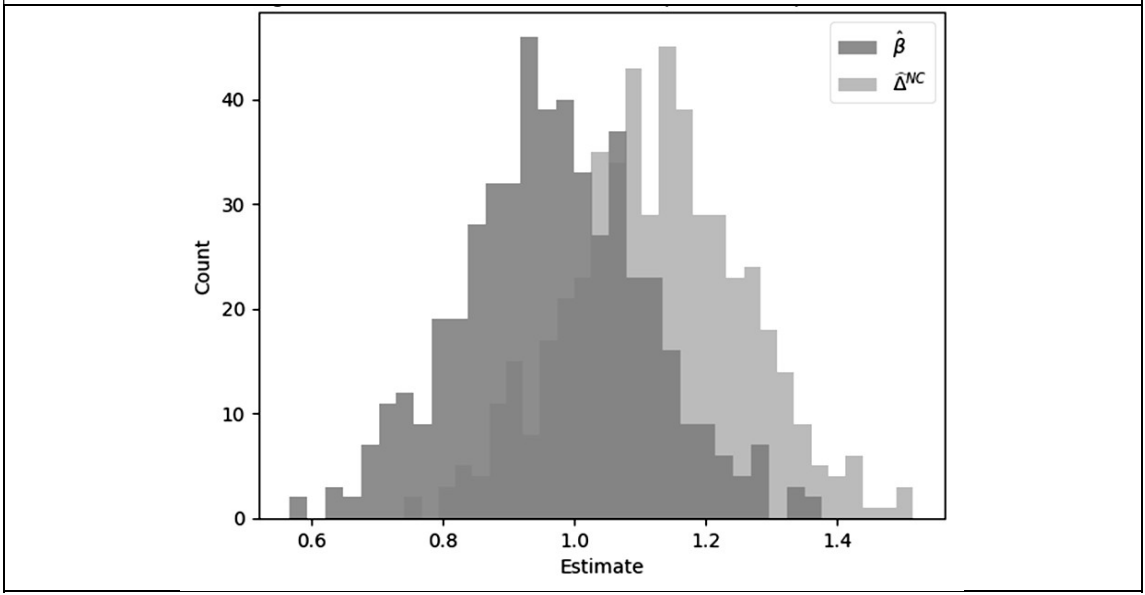

Figure 2. Monte Carlo distributions of the estimated direct effect and the implied equilibrium effect

Figure 2 makes this distinction explicit by comparing the Monte Carlo distributions of the estimated direct effect, $\hat{\beta}$, and the implied equilibrium effect, $\hat{\Delta}^{NC}$. Both objects are obtained from the same estimated spatial autoregressive model, using the same data and the same interaction structure. The marked difference between their distributions therefore does not reflect a change in estimator, specification, or identification strategy. It reflects a change in the counterfactual question being asked. Interpreting $\hat{\beta}$ as a causal effect corresponds to a partial-equilibrium counterfactual in which the outcomes of other units are held fixed. Mapping the same estimates through the spatial multiplier yields a network-consistent counterfactual that allows full equilibrium adjustment through the interaction structure. Much of the applied spatial literature implicitly adopts the latter interpretation when reporting direct and indirect impacts, often without making the underlying counterfactual explicit (Debarsy and Le Gallo, 2025). Figure 2 shows that this choice is not innocuous: even under exogenous treatment assignment and correct model specification, partial-equilibrium and equilibrium causal effects differ mechanically.

We finally relax the assumption of exogenous treatment assignment to illustrate how interdependence amplifies standard endogeneity concerns when equilibrium counterfactuals are considered. In this extension, treatment is correlated with unobserved productivity shocks, breaking the global exogeneity condition required for identification of network-consistent effects (Hudgens and Halloran, 2008). Table 4 reports Monte Carlo results under this confounded assignment.

Table 4. Monte Carlo performance under confounded assignment

| Estimator | True value | Mean estimate | Bias | RMSE | SD |
|---|---|---|---|---|---|
| $\hat{\beta}$ | 1.000 | 0.978 | −0.022 | 0.140 | 0.138 |
| $\hat{\rho}$ | 0.400 | 0.683 | +0.283 | 0.287 | 0.052 |
| $\widehat{\Delta}^{NC}$ | 1.037 | 1.124 | +0.086 | 0.157 | 0.131 |

As expected, the direct-effect estimate $\hat{\beta}$ is biased. In addition, the implied equilibrium effect $\widehat{\Delta}^{NC}$ displays substantially larger bias and dispersion. Network feedback magnifies the consequences of endogeneity when estimates are mapped into equilibrium counterfactuals. To quantify this amplification, we compute the ratio of biases between the equilibrium and direct-effect estimates. In the simulation, this ratio is approximately −3.9 in absolute value, indicating that equilibrium counterfactual distortions are several times larger than distortions in the direct-effect estimate. Figure 3 complements this result comparing the distribution of $\widehat{\Delta}^{NC}$ under exogenous and confounded assignment, highlighting that even mild departures from global exogeneity can lead to substantial distortions in equilibrium causal inference.

**Figure 3. Implied equilibrium effects: exogenous vs confounded assignment**

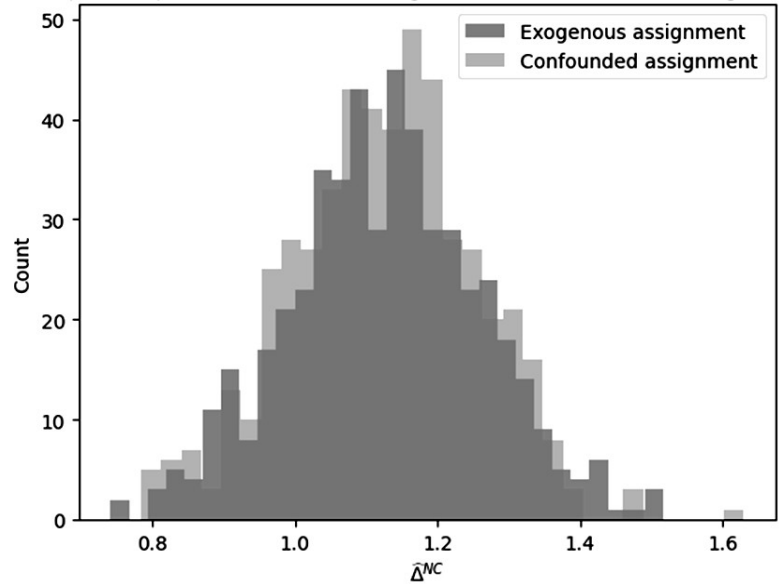

The Monte Carlo exercise delivers three clear lessons. First, causal effects under interdependence are not uniquely defined: partial-equilibrium and network-consistent effects differ mechanically, even before estimation. Second, correct estimation of a spatial model does not guarantee meaningful causal interpretation beyond the partial-equilibrium effect; equilibrium counterfactuals are distinct objects. Third, when treatment assignment is confounded, equilibrium causal effects are particularly fragile, as network feedback amplifies estimation distortions. Taken together, these results reinforce the central message of the paper. The contribution does not lie in proposing a more sophisticated Monte Carlo design or a novel estimator. Rather, it lies in clarifying that counterfactual interpretation, not econometric complexity, is the binding constraint for causal inference under interdependence. Learning interaction structures and estimating spatial models are necessary steps, but without explicit counterfactual definitions and corresponding identifying assumptions, causal conclusions remain fundamentally ambiguous.

## 7. Conclusion

This paper revisits a foundational question in empirical economics: what does it mean to estimate a causal effect when economic units interact? While a growing literature recognizes the ubiquity of spillovers, feedback, and network effects, much of empirical practice continues to interpret estimated parameters as causal without explicitly defining the counterfactuals that give those effects meaning. We argue that this gap is not merely technical, but conceptual. In environments characterized by endogenous spatial or network interdependence, causal effects are not uniquely defined objects. They are intrinsically tied to assumptions about how treatment-induced shocks propagate through the interaction structure.

A central contribution of the paper is to show that learning the interaction structure, while necessary, is not sufficient for causal interpretation. Recent advances that estimate networks or spatial weights from predetermined characteristics represent an important step toward realism. However, even with a correctly specified or consistently learned interaction structure—and even under ideal identifying conditions—causal conclusions remain ambiguous unless the researcher explicitly specifies the counterfactual regime governing adjustment. Without such a specification, estimated spillover parameters do not correspond to a well-defined economic object. We formalize this insight by distinguishing between three economically meaningful counterfactual regimes: partial-equilibrium, local-interaction, and network-consistent (equilibrium) counterfactuals. Each regime answers a different policy question, relies on different identifying assumptions, and maps standard spatial autoregressive estimates into distinct causal effects. By making these distinctions explicit, the framework clarifies what spatial and network models do—and do not—identify. We show that equilibrium causal effects require substantially stronger assumptions than direct or local effects, and that network feedback can mechanically amplify bias when those assumptions fail.

The Monte Carlo simulation deliberately adopts a simple and transparent design. This simplicity is intentional. The objective is not to introduce a sophisticated data-generating process or to assess the relative performance of competing estimators. Rather, it is to isolate the conceptual point that differences in causal conclusions arise from counterfactual interpretation, not from econometric complexity. Even in a canonical spatial autoregressive model with exogenous treatment assignment and known interaction structure, partial-equilibrium and equilibrium effects differ mechanically. When identifying assumptions are violated, equilibrium effects are particularly fragile, as feedback amplifies distortions. The simulation underscores that counterfactual ambiguity is present before estimation, not as a consequence of it.

More broadly, the paper contributes to ongoing efforts to reconcile modern causal inference with the reality of interconnected economic systems. Policies affecting firms, regions, or markets rarely operate in isolation. Evaluating such policies requires not only credible sources of variation, but also a clear statement of which economic responses are being allowed to occur in the counterfactual world. By placing counterfactual definitions at the center of causal analysis under interdependence, this paper provides a unifying perspective that complements recent critiques of spatial identification and the literature on interference. While the framework is developed in the context of linear spatial autoregressive models, its core message is not model-specific. The non-uniqueness of causal effects under interdependence extends naturally to nonlinear models, strategic interaction models, network games, and fully structural environments in which agents respond optimally to each other's actions. In all such settings, causal effects cannot be defined independently of assumptions about equilibrium adjustment and feedback. The SAR model serves here as a transparent and familiar vehicle for making this point precise.

Several directions for future research follow naturally. One is to extend the framework to environments in which treatments themselves are endogenously determined within the network. Another is to study heterogeneity in counterfactual effects across units with different network positions. A further avenue is to integrate the counterfactual perspective developed here with structural models, allowing researchers to discipline counterfactual regimes using economic theory while maintaining clarity about causal interpretation. The central lesson of the paper is that causal inference under interdependence is not primarily a problem of estimation, but of definition. Without explicit counterfactual assumptions, causal effects are ill-defined—even in well-specified and well-estimated models. Making these assumptions explicit is therefore not a technical detail, but a prerequisite for meaningful causal interpretation. We hope that this perspective will help bring greater clarity, discipline, and transparency to empirical work in spatial and networked economic environments.

## References

Acemoglu, D., Carvalho, V. M., Ozdaglar, A., and Tahbaz-Salehi, A. (2012). The network origins of aggregate fluctuations. *Econometrica*, *80*(5), 1977-2016.

Aghion, P., Dechezleprêtre, A., Hemous, D., Martin, R. and Van Reenen, J. (2016). Carbon taxes, path dependency, and directed technical change: Evidence from the auto industry. *Journal of Political Economy*, *124*(1), 1-51.

Anselin, Luc (1988). *Spatial econometrics: Methods and models.* Dordrecht: Kluwer Academic Publishers.

Athey, S., Eckles, D., and Imbens, G. W. (2018). Exact p-values for network interference. *Journal of the American Statistical Association*, *113*(521), 230-240.

Bramoullé, Y., Djebbari, H. and Fortin, B. (2009). Identification of peer effects through social networks. *Journal of econometrics*, *150*(1), 41-55.

Bramoullé, Yann, Habiba Djebbari, and Bernard Fortin. (2020). Peer Effects in Networks: A Survey. *Annual Review of Economics*, 12(1), 603–629.

Busso, M., Gregory, J., and Kline, P. (2013). Assessing the incidence and efficiency of a prominent place based policy. *American Economic Review*, *103*(2), 897-947.

Carvalho, V. M. (2014). From micro to macro via production networks. *Journal of Economic Perspectives*, *28*(4), 23-48.

Debarsy, N., and Le Gallo, J. (2025). Identification of Spatial Spillovers: Do's and Don'ts. *Journal of Economic Surveys*.

De Loecker, J., Eeckhout, J., and Unger, G. (2020). The rise of market power and the macroeconomic implications. *The Quarterly journal of economics*, *135*(2), 561-644.

Gao, M., and Ding, P. (2025). Causal inference in network experiments: regression-based analysis and design-based properties. *Journal of Econometrics*, *252*, 106119.

Gibbons, Stephen, and Henry G. Overman. (2012). Mostly Pointless Spatial Econometrics? *Journal of Regional Science*, 52(2), 172–191.

Heckman, James J., and Edward Vytlacil. (2007). Econometric Evaluation of Social Programs, Part II: Using the Marginal Treatment Effect to Organize the Evidence. In *Handbook of Econometrics*, Vol. 6B, ed. James J. Heckman and Edward E. Leamer. Amsterdam: Elsevier

Hudgens, M. G., and Halloran, M. E. (2008). Toward causal inference with interference. *Journal of the American Statistical Association*, *103*(482), 832-842.

Hudgens, Michael G., and M. Elizabeth Halloran. (2008). Toward Causal Inference with Interference. *Journal of the American Statistical Association*, 103(482), 832–842.

Imbens, Guido W. (2024). Causal Inference in the Social Sciences. *Annual Review of Statistics and Its Application*, 11, 123–152.

Imbens, G. W., and Rubin, D. B. (2015). *Causal inference in statistics, social, and biomedical sciences*. Cambridge university press.

Kline, P., and Moretti, E. (2014). People, places, and public policy: Some simple welfare economics of local economic development programs. *Annual Review of Economics*, *6*(1), 629-662.

LeSage, James P., and R. Kelley Pace (2009). *Introduction to spatial econometrics.* Boca Raton, FL: CRC Press.

Leung, Michael P. (2022). Causal Inference Under Approximate Neighborhood Interference. *Econometrica*, 90(1), 267–293.

Manski, C. F. (1993). Identification of endogenous social effects: The reflection problem. *The review of economic studies*, *60*(3), 531-542.

Merk, N., and Otto, P. (2022). Estimation of the spatial weighting matrix for regular lattice data—An adaptive lasso approach with cross-sectional resampling. *Environmetrics*, 33(3), e2736.

Qiu, Yixiao, and Li Tong (2021). Causal inference with spatial spillovers. *Journal of Econometrics*, 225(2): 401–421.

Rubin, D. B. (1974). Characterizing the estimation of parameters in incomplete-data problems. *Journal of the American Statistical Association*, *69*(346), 467-474.

Sävje, F., Aronow, P., and Hudgens, M. (2021). Average treatment effects in the presence of unknown interference. *Annals of statistics*, *49*(2), 673.

Sävje, Fredrik, Peter M. Aronow, and Michael G. Hudgens. (2021). "Average Treatment Effects in the Presence of Unknown Interference." *The Annals of Statistics*, 49(2), 673–701.

**Appendix. Identification under Alternative Counterfactual Regimes**

**Proof Proposition 1**

Under Assumption 1 and conditional exogeneity $D_i \perp \varepsilon_i \mid X_i$, the partial-equilibrium causal effect $\Delta_i^{PE}$ is identified by $\beta$.

**Proof**

The outcome system satisfies the SAR equilibrium condition

$$Y = \rho W Y + \beta D + X \gamma + \varepsilon \tag{A1}$$

Let $Y^0$ denote the baseline outcome vector under the reference assignment, i.e., the outcomes defining the pretreatment environment used in the partial equilibrium counterfactual. The partial-equilibrium counterfactual changes $D_i$, while holding the outcomes of all other companies fixed at their baseline levels $Y_{-i}^0$. Formally,

$$\Delta_i^{PE} = Y_i(D_i = 1, Y_{-i} = Y_{-i}^0) - Y_i(D_i = 0, Y_{-i} = Y_{-i}^0) \tag{A2}$$

Under this regime, the interaction term entering firm $i$'s outcome equation is fixed. Define

$$m_i \equiv (WY^0)_i = \sum_{j \neq i} wij Y_j^0 \tag{A3}$$

which is constant with respect to $D_i$ by construction of the counterfactual since $Y_{-i}$ is held fixed at $Y_{-i}^0$. The i-th equation implied by (A1) under the partial equilibrium regime is therefore

$$Y_i = \rho m_i + \beta D_i + X_i^\top \gamma + \varepsilon_i \tag{A4}$$

Taking conditional expectations given $X_i$ and using conditional exogeneity $D_i \perp \varepsilon_i \mid X_i$ implies

$$\mathbb{E}[Y_i \mid D_i = 1, X_i] - \mathbb{E}[Y_i \mid D_i = 0, X_i] \tag{A5}$$
$$= \beta + \underbrace{\mathbb{E}[\varepsilon_i \mid D_i = 1, X_i] - \mathbb{E}[\varepsilon_i \mid D_i = 0, X_i]}_{=0} = \beta$$

since the constant terms $\rho m_i$ and $X_i^T \gamma$ cancel and the error difference is zero by conditional exogeneity. Hence $\Delta_i^{PE} = \beta$, establishing that $\beta$ admits a causal interpretation as the partial-equilibrium effect.

**Proof of Proposition 2 (Local-Interaction Effects)**

We consider outcomes satisfying the equilibrium spatial autoregressive condition:

$$Y = \rho W Y + \beta D + X\gamma + \varepsilon \qquad (A6)$$

where the interaction matrix $W = W(\theta; X)$ is predetermined with respect to treatment assignment, as stated in Assumption 1. Let $e_i$ denote the i-th canonical basis vector.

The local-interaction counterfactual allows a change in $D_i$ to affect outcomes through direct effects and first order neighbor responses, while excluding higher-order feedback effects. Define the one-step local response mapping:

$$Y^{LI}(D) = (I + \rho W)(\beta D + X\gamma + \varepsilon) \qquad (A7)$$

which corresponds to the first order expansion of the full equilibrium mapping $(I - \rho W)^{-1}$. This term captures the idea that neighbors respond once to the treatment, but that their responses do not trigger further rounds of adjustment.

The local-interaction effect of treating firm $i$ holding $D_{-i}$ fixed is defined as

$$\Delta_i^{LI} = Y_i^{LI}(D_i = 1, D_{-i}) - Y_i^{LI}(D_i = 0, D_{-i}) \qquad (A8)$$

Substituting the local response mapping (A7) yields

$$\Delta_i^{LI} = e_i^T (I + \rho W)\beta (D^{(1)} - D^{(0)}) \qquad (A9)$$

where $D^{(1)} = (D_i = 1, D_{-i})$ and $D^{(0)} = (D_i = 0, D_{-i})$. Since $D^{(1)} - D^{(0)} = e_i$, we obtain

$$\Delta_i^{LI} = \beta e_i^T (I + \rho W) e_i = \beta \qquad (A10)$$

using the fact that $W_{ii} = 0$.

Thus, allowing one-step neighbor responses does not alter the treated firms' own effect relative to partial-equilibrium. The local-interaction counterfactual becomes informative for spillovers. For any firm $j$ such that $W_{ji} > 0$, the causal spillover from treating firm i to firm j under the local-interaction counterfactual is

$$\Delta^{LI}_{j \leftarrow i} = Y_j^{LI}(D^{(1)}) - Y_j^{LI}(D^{(0)}) = e_j(I + \rho W)\beta e_i = \beta \rho W_{ji} \quad (A11)$$

Identification follows from the conditional mean difference.

$$\mathbb{E}[Y_j \mid D_i = 1, X] - \mathbb{E}[Y_j \mid D_i = 0, X] \quad (A12)$$

Under local exogeneity (Assumption 2),

$$D_i \perp \varepsilon_j \mid X \text{ for all j such that } W_{ji} > 0 \quad (A13)$$

variation in $D_i$ affects $Y_j$ only through the local-response channel. Hence, (A12) recovers $\Delta^{LI}_{j \leftarrow i}$.

## Proof of Proposition 3 (Network-Consistent Effects)

**Proof.**

Outcomes satisfy the spatial autoregressive equilibrium condition

$$Y = \rho WY + \beta D + X\gamma + \quad (A14)$$

where the interaction matrix $W = W(\theta; X)$ is predetermined with respect to treatment assignment (Assumption 1). Solving for outcomes yields the reduced-form equilibrium mapping

$$Y(D) = (I - \rho W)^{-1}(\beta D + X\gamma + \varepsilon) \quad (A15)$$

The network-consistent counterfactual allows the treatment-induced shock to propagate fully through the interaction structure, including all higher-order feedback effects, until a new equilibrium is reached. The causal effect of treating firm i, holding all other treatments fixed at zero, is defined as

$$\Delta^{NC}_i = Y_i(D_i = 1, D_{-i} = 0) - Y_i(D_i = 0, D_{-i} = 0). \quad (A16)$$

Substituting the equilibrium mapping yields

$$\Delta_i^{NC} = e_i^\top (I - \rho W)^{-1} \beta e_i. \tag{A17}$$

where $e_i$ is the i-th canonical basis vector. This expression captures the total effect of the treatment on firm i, including all indirect and feedback effects transmitted through the interaction network.

To establish identification from observed data, consider the conditional mean difference

$$\mathbb{E}[Y_i \mid D_i = 1, X] - \mathbb{E}[Y_i \mid D_i = 0, X]. \tag{A18}$$

Under global exogeneity (Assumption 3),

$$D \perp \varepsilon \mid X. \tag{A19}$$

the entire treatment assignment vector is conditionally independent of the vector of unobserved shocks. This condition ensures that variation in $D_i$ is uncorrelated not only with $\varepsilon_i$, but also with all shocks affecting other firms whose outcomes enter $Y_i$ through the equilibrium feedback embodied in $(I - \rho W)^{-1}$. As a result, the conditional mean difference in (A18) recovers the equilibrium contrast $\Delta_i^{NC}$. Hence, under Assumptions 1 and 3, the network-consistent causal effect is identified by the equilibrium mapping implied by the spatial autoregressive model.